\documentclass[journal=ancac3,manuscript=article]{achemso}

\usepackage[version=3]{mhchem} % Formula subscripts using \ce{}

\newcommand*{\3}{$^3$He}

 \usepackage{siunitx}
 \usepackage{amssymb}
 \usepackage{hyperref}
\hypersetup{
    colorlinks=false,
    linkcolor=blue,
    filecolor=magenta,      
    urlcolor=cyan,
    pdftitle={NbTi nanowire oscillators},
    pdfpagemode=FullScreen,
    }

\author{Samuli Autti}
\affiliation[Lancaster University]{Physics Department, Lancaster University, Lancaster, LA1 4YB, UK}
\email{s.autti@lancaster.ac.uk}

\author{Andrew Casey}
\affiliation[Royal Holloway University of London]{Department of Physics, Royal Holloway University of London, Egham, Surrey, TW20 0EX, UK}

\author{Marie Connelly}
\affiliation[Lancaster University]{Physics Department, Lancaster University, Lancaster, LA1 4YB, UK}

\author{Neda Darvishi}
\affiliation[Royal Holloway University of London]{Department of Physics, Royal Holloway University of London, Egham, Surrey, TW20 0EX, UK}

\author{Paolo Franchini}
\affiliation[Royal Holloway University of London]{Department of Physics, Royal Holloway University of London, Egham, Surrey, TW20 0EX, UK}
\alsoaffiliation[Lancaster University]{Physics Department, Lancaster University, Lancaster, LA1 4YB, UK}

\author{James Gorman}
\affiliation[Lancaster University]{Physics Department, Lancaster University, Lancaster, LA1 4YB, UK}

\author{Richard P. Haley}
\affiliation[Lancaster University]{Physics Department, Lancaster University, Lancaster, LA1 4YB, UK}

\author{Petri J. Heikkinen}
\affiliation[Royal Holloway University of London]{Department of Physics, Royal Holloway University of London, Egham, Surrey, TW20 0EX, UK}

\author{Ashlea Kemp}
\affiliation[Royal Holloway University of London]{Department of Physics, Royal Holloway University of London, Egham, Surrey, TW20 0EX, UK}

\author{Elizabeth Leason}
\affiliation[Royal Holloway University of London]{Department of Physics, Royal Holloway University of London, Egham, Surrey, TW20 0EX, UK}

\author{John March-Russell}
\affiliation[University of Oxford]{Rudolf Peierls Centre for Theoretical Physics, University of Oxford, 1 Keble Road, Oxford, OX1 3NP, UK}

\author{Jocelyn Monroe}
\affiliation[Royal Holloway University of London]{Department of Physics, Royal Holloway University of London, Egham, Surrey, TW20 0EX, UK}

\author{Theo Noble}
\affiliation[Lancaster University]{Physics Department, Lancaster University, Lancaster, LA1 4YB, UK}

\author{George R. Pickett}
\affiliation[Lancaster University]{Physics Department, Lancaster University, Lancaster, LA1 4YB, UK}

\author{Jonathan R. Prance}
\affiliation[Lancaster University]{Physics Department, Lancaster University, Lancaster, LA1 4YB, UK}

\author{Xavier Rojas}
\affiliation[Royal Holloway University of London]{Department of Physics, Royal Holloway University of London, Egham, Surrey, TW20 0EX, UK}

\author{Tineke Salmon}
\affiliation[Lancaster University]{Physics Department, Lancaster University, Lancaster, LA1 4YB, UK}

\author{John Saunders}
\affiliation[Royal Holloway University of London]{Department of Physics, Royal Holloway University of London, Egham, Surrey, TW20 0EX, UK}

\author{Jack Slater}
\affiliation[Lancaster University]{Physics Department, Lancaster University, Lancaster, LA1 4YB, UK}

\author{Robert Smith}
\affiliation[Royal Holloway University of London]{Department of Physics, Royal Holloway University of London, Egham, Surrey, TW20 0EX, UK}

\author{Michael D. Thompson}
\affiliation[Lancaster University]{Physics Department, Lancaster University, Lancaster, LA1 4YB, UK}

\author{Stephen M. West}
\affiliation[Royal Holloway University of London]{Department of Physics, Royal Holloway University of London, Egham, Surrey, TW20 0EX, UK}

\author{Luke Whitehead}
\affiliation[Lancaster University]{Physics Department, Lancaster University, Lancaster, LA1 4YB, UK}

\author{Vladislav V. Zavjalov}
\affiliation[Lancaster University]{Physics Department, Lancaster University, Lancaster, LA1 4YB, UK}

\author{Kuang Zhang}
\affiliation[University of Sussex]{Department of Physics and Astronomy, University of Sussex, Brighton, BN1 9QH, UK}

\author{Dmitry E. Zmeev}
\affiliation[Lancaster University]{Physics Department, Lancaster University, Lancaster, LA1 4YB, UK}
\email{d.zmeev@lancaster.ac.uk}

\title[NbTi nanowire oscillators]{Long nanomechanical resonators with circular cross-section}

\begin{document}

\abstract{Fabrication of superconducting nanomechanical resonators for quantum research, detectors and devices traditionally relies on a lithographic process, resulting in oscillators with sharp edges and a suspended length limited to a few \SI{100}{\micro\meter}. We report a low-investment top-down approach to fabricating NbTi nanowire resonators with suspended lengths up to several millimetres and diameters down to \SI{100}{\nano\meter}. The nanowires possess high critical currents and fields, making them a natural choice for magnetomotive actuation and sensing. This fabrication technique is independent of the substrate material, dimensions and layout and can readily be adapted to fabricate nanowire resonators from any metal or alloy with suitable ductility and yield strength. Our work thus opens access to a new class of nanomechanical devices with applications including microscopic and mesoscopic investigations of quantum fluids, detecting dark matter and fundamental materials research in one-dimensional superconductors in vacuum.}

%possibly useful resources https://pubs.acs.org/doi/abs/10.1021/acs.nanolett.0c01027
% https://www.nature.com/articles/nnano.2009.267
% https://www.nature.com/articles/s41565-019-0605-9
% vijayan2023scalable https://www.nature.com/articles/s41565-022-01254-6

\maketitle

\clearpage
%\section{Main}

Miniaturised mechanical oscillators are ubiquitous tools for sensing applications \cite{ghaemi2022} and for fundamental quantum research \cite{leggett2002,collin2022mesoscopic}. The QUEST-DMC collaboration\cite{QUEST}  is constructing a superfluid $^3$He based dark matter detector, and the feasibility of the projected detection scheme \cite{QUEST} depends on recording the temperature of the superfluid using a new class of superconducting nanomechanical resonators. The resonators should be millimetres long to produce a large signal and a low resonance frequency and come with no sharp edges that would interfere invasively with the superfluid. Equally, superfluid $^3$He at length scales smaller than 100~nm is relatively unexplored mechanically. For example, the quantum-classical interface, which remains one of the major open problems in modern physics, is directly experimentally accessible at the edges of this superfluid system\cite{Autti2023,autti2020fundamental} provided mechanical instruments thin enough to directly fit inside it ($\sim$100\,nm). We expect to find, for example, Majorana fermions there \cite{Nagai2008,Murakawa2011, Sauls2013}. In this article we develop a fabrication process that allows making millimetres long, round superconducting nanowire resonators down to 100\,nm in diameter, thus enabling these and many other experiments.

Nanomechanical oscillators \cite{Bachtold2022,Wei2021} are typically fabricated in either a top-down process using lithography and etching, see for example \cite{Golokolenov22,kamppinen2019nanomechanical}, or in a bottom-up process based on growing the desired structures in place \cite{xu2022nanomechanical,tavernarakis2018optomechanics,laird2015quantum}. These techniques are tied to a particular fabrication process flow which limits the selection of possible substrate and resonator materials, geometries and arrangement. In this Article we demonstrate a low-investment technique for making NbTi nanowires using a set of wire drawing dies, followed by manually assembling the nanowire resonators on an arbitrary device platform using tweezers and an optical microscope. We show that standard wire bonding techniques can be used to make low-resistance contacts to the resonator, and that tension in the resonator can be manipulated during the assembly down to the strain-free resonator limit \cite{zhou2020approaching,Bachtold2022}. The suspended length of these resonators can be several millimetres and the diameter can be lowered to \SI{100}{\nano\metre}.

\section{Nanowire resonators}
\begin{figure}%
\centering
\includegraphics[width=\linewidth]{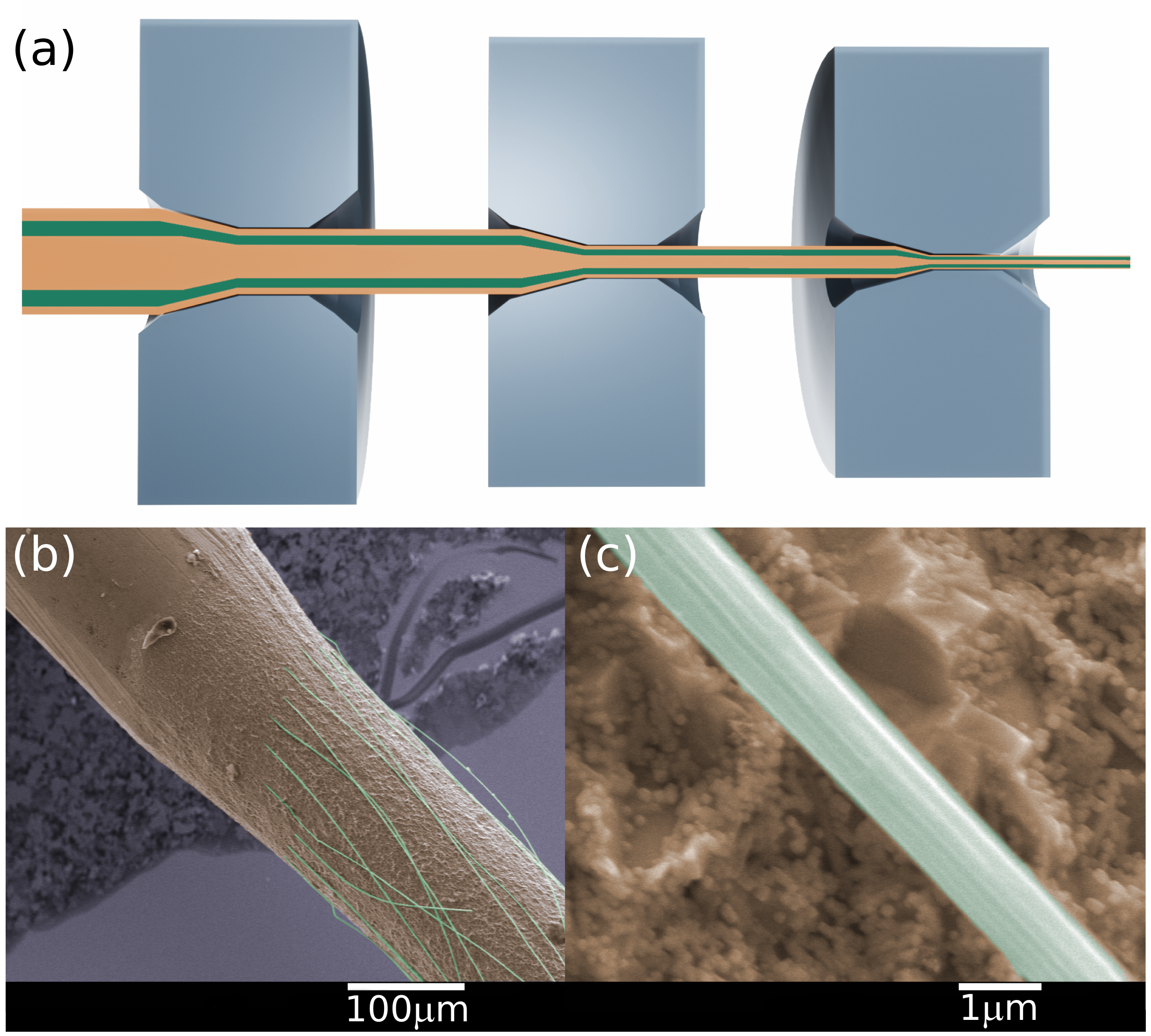}\caption{{\bf Nanowire fabrication:} {\bf (a)} A cable with NbTi filaments embedded in a copper matrix is pulled through a series of drawing dies, reducing the diameter of both the copper wire and the NbTi filaments; {\bf (b)} False colour SEM image showing the NbTi filaments exposed after partial etching of the copper matrix. Scale bar is \SI{100}{\micro\metre}; {\bf (c)} False colour SEM image showing a close up of one of the filaments prior to being thinned, demonstrating the initial diameter to be \SI{1}{\micro\metre}. Scale bar is \SI{1}{\micro\metre}. }\label{fig:Nanowire_Fabrication}
\end{figure}

The fabrication process starts with a \SI{200}{\micro\metre} diameter copper wire containing twenty \SI{1}{\micro\metre}-thick round NbTi filaments (see Figure \ref{fig:Nanowire_Fabrication}) that follow the perimeter of the cable. The cable is drawn through a set of diamond wire dies manufactured by the Esteves Group (\href{http://www.estevesgroup.com}{estevesgroup.com}) to reduce the diameter of the cable, and the filaments are reduced proportionally. The micrographs in Fig.~\ref{fig:Nanowire_Fabrication} show the original cable and the NbTi filaments after etching with concentrated nitric acid.

\begin{figure*}%
\centering
\includegraphics[width=\linewidth]{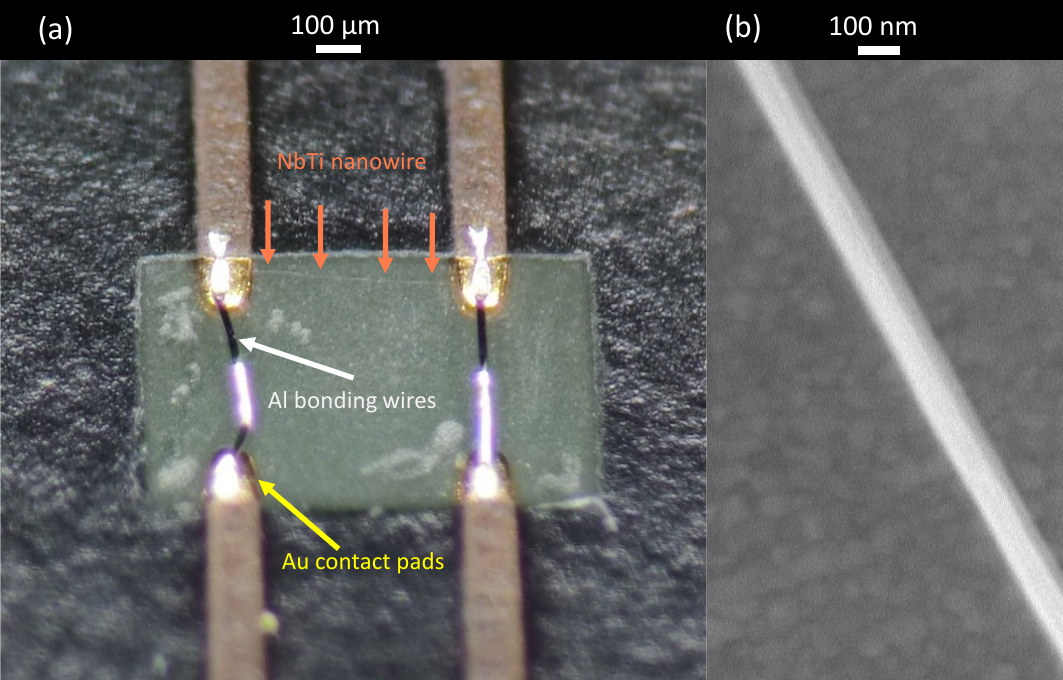}
\caption{{\bf Images of a 100 nm diameter nanowire resonator:} {\bf (a)} An optical image showing a 4-point measurement circuit. The nanowire (indicated by orange arrows) is suspended between the gold-plated contact pads (yellow arrow) and bonded to them with aluminium wires (white arrow). The pads are elevated \SI{40}{\micro\metre} above the surface. Scale bar is \SI{100}{\micro\metre} and the rectangular hole in the PCB solder-mask has a width of \SI{1}{mm}. In the characterisation experiments shown in Fig.~\ref{fig:100nm_resonance}, the magnetic field is applied perpendicular to the plane of the circuit board shown here. {\bf (b)} SEM image of a nanowire, from the same cable as the wire shown in (a), on a gold substrate where the diameter is found to be around \SI{100}{\nano\metre}. This nanowire was fabricated from a \SI{20}{\micro\metre} cable, i.e. the cable shown in Fig.~\ref{fig:Nanowire_Fabrication}(b) drawn down to a tenth of its initial diameter. The filament can be seen to have shrunk in proportion. Scale bar is \SI{100}{\nano\metre}.}
\label{fig:100nm_wire}
\end{figure*}

In order to produce individual nanowires, the copper cable is etched until the NbTi filaments are exposed. The etching can be stopped by diluting the acid with water and, crucially, the filaments are not subjected to surface tension of the liquid at this point. We can now select one filament with tweezers under an optical stereo microscope, and remove it from the solution by pulling it out vertically. Figure \ref{fig:100nm_wire} shows a \SI{100}{\nano\metre} NbTi nanowire that has been placed between two gold pads on a standard circuit board using tweezers. The pads are separated by \SI{600}{\micro\metre}. The nanowire is suspended between the elevated pads where van der Waals forces are sufficient to hold the wire in place. The wire is then fixed by bonding Al wires onto the pads, trapping the NbTi filament between the gold pads and the bond wire. In this way the nanowire can be easily integrated in a measurement circuit. We emphasise that this fabrication process allows placing the nanowire on any platform so long as sufficient space and an appropriate substrate for wire bonding are provided.

The resulting suspended nanowires make versatile nanoelectromechanical resonators (NEMS). Passing an alternating current through the wire in a magnetic field results in a Lorentz force driving the NEMS in a direction transverse to the magnetic field. In Fig.~\ref{fig:100nm_wire}, the magnetic field is oriented perpendicular to the plane of the circuit. Thus, the resonator is driven to move in the plane of the circuit board. The second pair of contact pads is used to measure the induced Faraday voltage across the resonator, which directly measures the resonator velocity in the plane of the motion.

\begin{figure}%
\centering
\includegraphics[width=\linewidth]{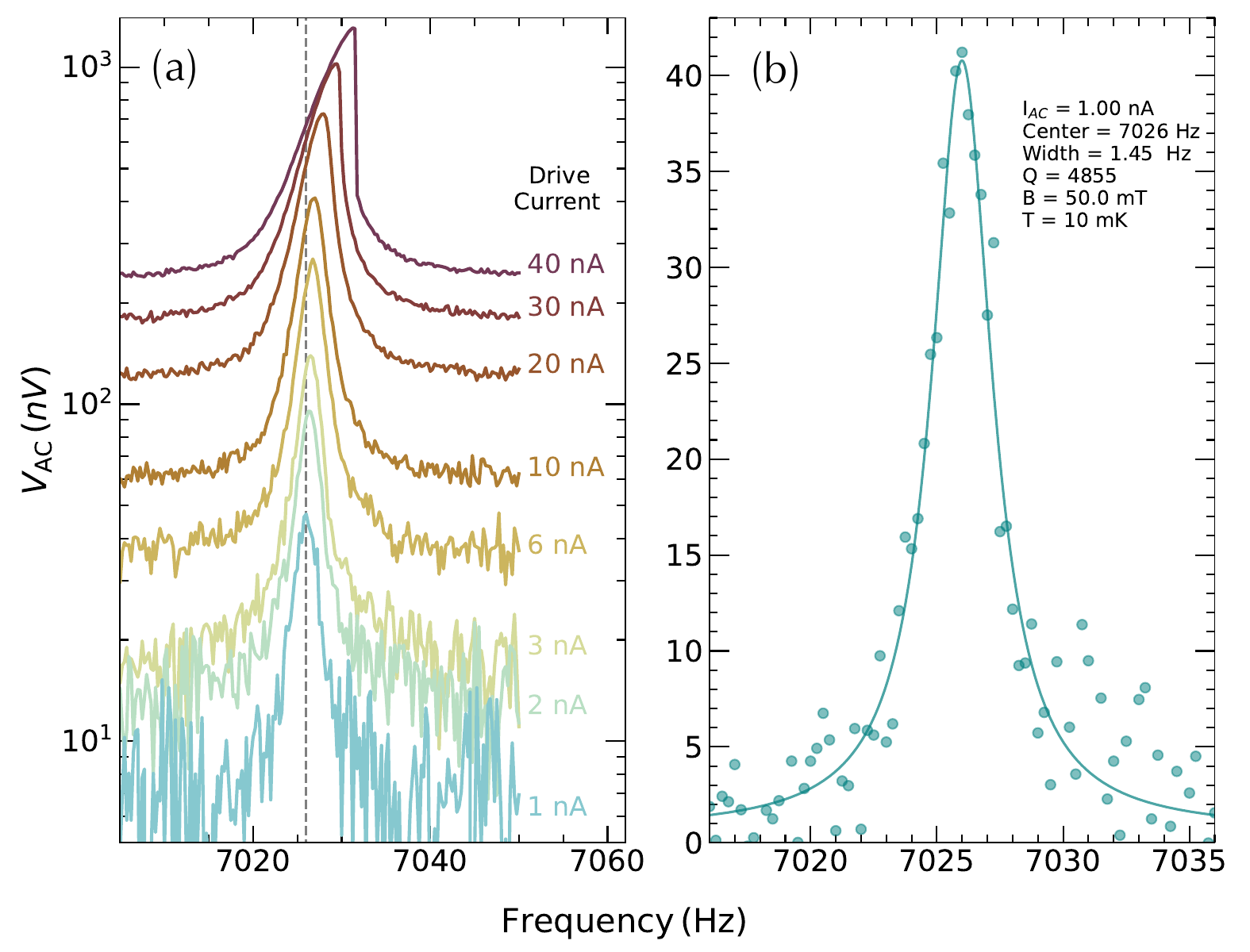}
\caption{{\bf Vacuum characterisation of the \SI{100}{\nano\metre} diameter, \SI{600}{\micro\metre} long vibrating wire resonator shown in Fig.~\ref{fig:100nm_wire}:} {\bf (a)} The in-phase voltage across the nanowire resonator at varying excitation currents shows the onset of non-linearity above \SI{3}{\nano\ampere}. Resonant curves measured at different drives are shifted vertically due to the non-zero contact resistance at \SI{50}{\milli\tesla}. {\bf (b)} Lorentzian fitting to the \SI{1}{\nano\ampere} resonance peak. The peak Faraday voltage corresponds to the maximum velocity of \SI{3.0}{\milli\metre\,\second^{-1}}. Measurements presented in both panels were carried out in 50\,mT magnetic field at 10\,mK temperature.} 
\label{fig:100nm_resonance}
\end{figure}

To work as an instrument for quantum fluids microscopy, we estimate the NEMS motion needs to remain linear up to a peak velocity of several \SI{}{\milli\metre\,\second^{-1}}. The NEMS also needs to be able to reach much higher velocities so that operation is not limited by applicable drive when the motion is being slowed down by a drag force from collisions with quasiparticle excitations in the fluid. To demonstrate resonator operation, we characterize the \SI{100}{\nano\metre} wire shown in Fig.~\ref{fig:100nm_wire} in vacuum. At 1\,nA drive in 50\,mT magnetic field the response is Lorentzian (linear) with amplitude of \SI{40}{\nano\volt}, a central frequency $f_0=7$\,kHz and quality factor 5000, as shown by the fit in Fig.~\ref{fig:100nm_resonance}b. This corresponds to a peak velocity of \SI{3}{\milli\metre \second^{-1}} on resonance. At velocities above \SI{15}{\milli\metre \second^{-1}} the resonance becomes nonlinear (Fig.~\ref{fig:100nm_resonance}a).

We have also fabricated and operated a $\SI{200}{\nano\metre}$ diameter resonator $\SI{3}{\milli\metre}$ in length with an aspect ratio of $1.5 \times 10^4$. This wire was moving freely in the ambient airflow, demonstrating absence of tension. Tension can be reapplied by using an adhesive with suitable surface tension to remove the slack from the wire. We believe at least \SI{1}{\centi\metre} suspended length can be obtained, as the nanowires are often longer than this before a part of the nanowire is bonded to make a resonator. The readout signal amplitude is proportional to the length of the resonator.

\begin{figure}%
\centering
\includegraphics[width=0.9\linewidth]{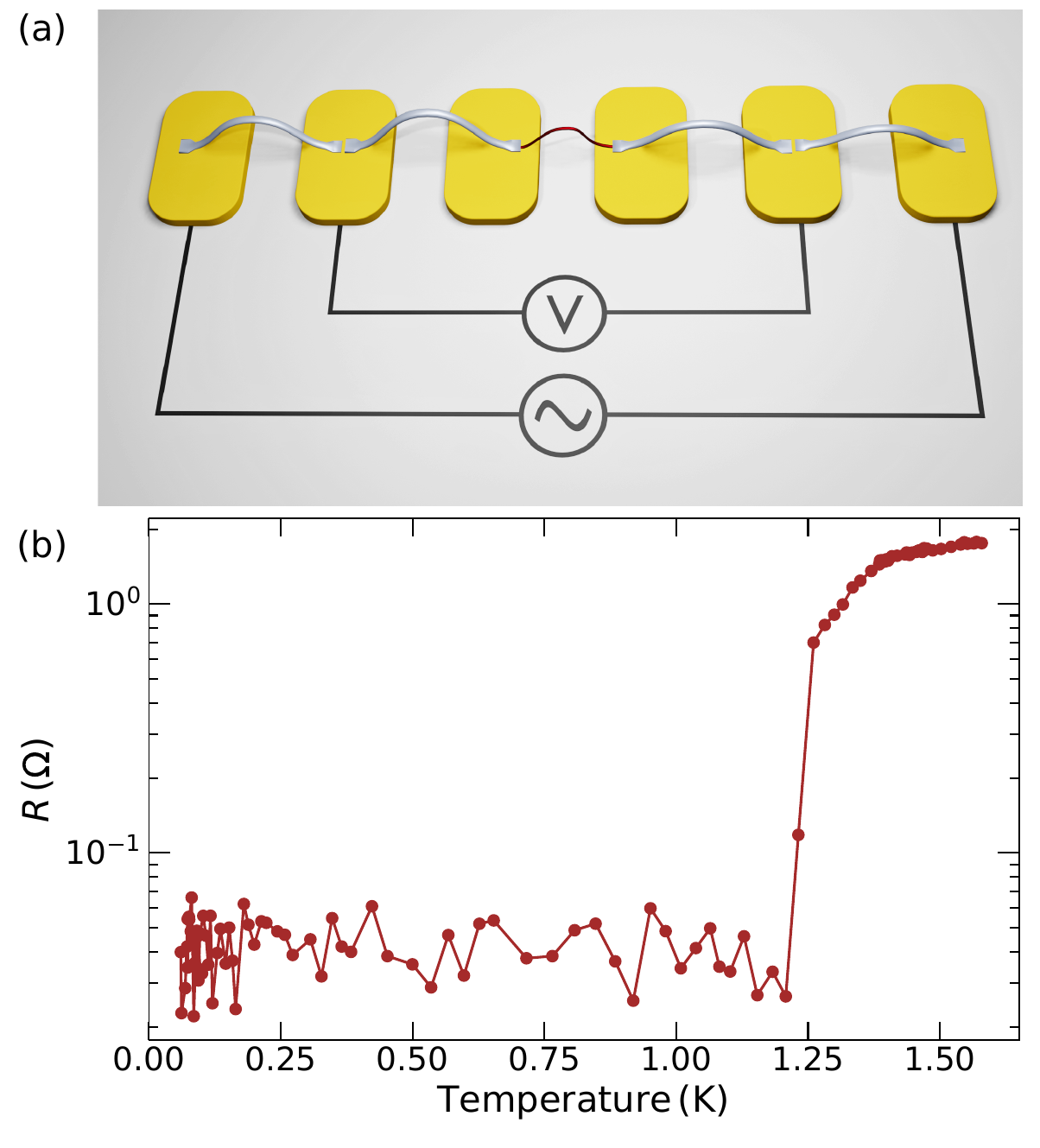}
\caption{{\bf Residual contact resistance between the nanowire and the aluminium bond wires:} {\bf (a)} A four-terminal resistance measurement is used to measure the contact resistance. The contacts between the nanowire (red) and the aluminium bond wires (grey) are in series with the nanowire. {\bf (b)} Resistance measured as a function of temperature with $\SI{50}{\nano\ampere}$ excitation at $\SI{73}\hertz$, far from resonance. Below 1.2 K the measurement is limited by the noise floor of the lock-in amplifier and sets an upper limit of the contact resistance to 30 m$\Omega$ per contact; increasing the drive to 700 nA we find a non-zero voltage which gives a resistance of 15 m$\Omega$ per contact.}
\label{fig:Residual}
\end{figure}

%\section{Electrical contacts to nanowires}

Superconductivity is an essential requirement for magneto-motive sensing at the lowest temperatures. For example, reaching the quantum limit of only a few phonons in a nanomechanical resonator\cite{cattiaux2021macroscopic} requires ultimate cooling and is therefore prohibited in the presence of Joule heating from readout. If the nanoresonator is immersed in a quantum fluid, Joule heating dissipated in the body of NEMS would result in local overheating of the surrounding liquid \cite{rickinson2020}. Typical superconducting NEMS devices are fabricated using vapour-deposited aluminium~\cite{Bradley2017,Midlik2022, Golokolenov22, kamppinen2019nanomechanical}. Aluminium has a relatively low critical field of about $\SI{10}{\milli\tesla}$.

Our nanowires superconduct up to the largest field we can apply in our characterisation setup, $\SI{8}{T}$, and we measure the critical temperature to be $\approx8$\,K. We have also measured the critical current in two different nanowires, 1000\,nm and 350\,nm in diameter, by increasing the current at 10\,mK temperature until the wire destructs through Joule heating in the normal state. The obtained critical current densities are \SI{10}{\kilo\ampere\, \milli\metre^{-2}} and \SI{15}{\kilo\ampere\, \milli\metre^{-2}}, respectively. The critical current density thus increases as the wire diameter decreases but remains similar to the values observed in bulk  wires~\cite{lee1987development,Green89,boutboul2006}. 

Ideally, superconducting resonators would be operated with a fully superconducting measurement circuit with superconducting contacts to the nanowire. We have measured the contact resistance between a nanowire and the Al bond wires at zero magnetic field such that the bond wires are superconducting. The 4-point measurement circuit (Fig.~\ref{fig:Residual}a) is designed so that the nanowire, bond wires and the contact resistance between them are all in series. Figure \ref{fig:Residual} shows the measured contact resistance as a function of temperature where we see that the aluminium bond wires go normal at 1.2\,K as expected. Below this temperature the resistance is lower than we can resolve with a 50\,nA excitation. Increasing the drive to 700\,nA, we find a non-zero voltage greater than the noise floor which gives an upper limit of $\SI{15}{\milli\ohm}$ per contact. It is possible that the contact between the Al bond wire and the NbTi nanowire is a tunnel junction and therefore superconducting at low excitation. Regardless, nanowire oscillators typically require  $\lesssim\SI{1}{\micro\ampere}$ drive, resulting in femtowatt level Joule heating if the total resistance is $\SI{30}{\milli\ohm}$. This level of dissipation is easily carried away by the substrate.

\section{Discussion}

We have demonstrated a technique for fabricating circular cross-section NbTi nanowire oscillators with diameters down to 100\,nm and aspect ratios over $10^4$, yielding untensioned resonance frequencies in the kilohertz range. The nanowires are robust in handling and can be hand-placed on an arbitrary platform, for example spanning across a volume far from any walls. We used NbTi nanowires possessing excellent superconducting properties, but the fabrication process can be applied to any metal or alloy with suitable properties and the resonator can be placed on any desired platform. Low resistance contacts with the nanowire resonator can be made simply by bonding a superconducting bond wire over the nanowire against an electric contact, but any existing nanowire actuation and sensing method \cite{Bachtold2022,Wei2021} can be reasonably adapted. The diamond die used by us to fabricate the smallest, $\SI{100}{\nano\metre}$, wire had the orifice diameter of $\SI{20}{\micro\metre}$. At the time of writing, $\SI{7}{\micro\metre}$ diameter dies were commercially available from Esteves group, enabling sub-$\SI{40}{\nano\metre}$ diameter NbTi resonators.

In the future, niobium bonding wires can be used to make superconducting contacts that work at high magnetic fields \cite{Jaszczuk1991}.

This new class of nanomechanical devices opens exciting research opportunities. We can place a nanowire resonator between a stationary and a mobile, piezo-actuated platform equipped with contact pads. This would allow adjusting tension in the resonator continuously and therefore controlling the wire's resonance frequency up to the typical frequencies of pre-tensioned resonators at several MHz. Another possibility is bridging a gap on an on-chip magnetic cooldown platform \cite{autti2022thermal}, also making a primary thermometer, and studying phonons in the resonator directly cooled to microkelvin temperatures. This may allow reaching the few-phonon limit in a resonator operated at MHz frequencies.  

In quantum fluids research, access to customisable high-quality superconducting nanomechanical instruments provides a broad range of opportunities. A $\SI{100}{\nano\metre}$ wire can be used to study the micro-structure of the superfluid coherence in \3 bulk and near walls \cite{Autti2023,Sauls2013,autti2020fundamental}, with transport experiments possibly revealing bound Majorana fermions \cite{Nagato19981135,Nagai2008,Murakawa2011}. Thermal fluctuations of the superfluid state remain unexplored and can be directly observed with a sensitive-enough mechanical probe \cite{Bradley2000rf} (see Methods for details on superfluid thermometry): combined with a quantum amplifier such as a SQUID, a nanowire could measure the shot noise of inbound collisions with thermal quasiparticle excitations down to the zero-temperature limit.  

Finally, long superconducting nanoresonators can be used to build a \3-based dark matter detector with unprecedented energy sensitivity for sub-eV dark matter candidates \cite{QUEST,Bradley1995}.

\section*{Methods}
\subsection*{Nanowire fabrication}
We start with a custom-made $\SI{0.2}{\milli\meter}$ cable with 20 NbTi filaments, $\SI{1}{\micro\meter}$ in diameter each. The cable was made by `Outokumpu Superconductors', currently known as Luvata Oy (\href{http://www.luvata.com}{luvata.com}). We draw the cable through  series of custom-made dies manufactured by the Esteves Group (\href{http://www.estevesgroup.com}{estevesgroup.com}) to reduce the diameter of the cable to the desired size. SEM imaging shows that the diameter of the filaments is reduced in proportion to the diameter of the cable as shown in Figs.~\ref{fig:Nanowire_Fabrication} and \ref{fig:100nm_wire}. We take a desired length of the resulting cable and partially etch the copper matrix using concentrated nitric acid. After etching is finished, we add plenty of water. Care should be taken to ensure that the filaments stay inside liquid at all times as the filaments can be easily broken by surface tension of water. We carefully manoeuvre a pair of sharp-tip tweezers (Ideal-tek, model 5SG.CX.0) under an optical microscope to separate out the NbTi filaments and to pull them out of the water. For separating the thinnest filaments it is essential that a bright blue light source is used. The nanowires are then transferred to the desired location using the same tweezers and laid down on contact pads.

\subsection*{Bonding} 
The nanowires are bonded using \SI{25}{\micro\metre} Al bond wire using a K\&S 4123 Manual Wedge Bonder. The ends of the nanowire are situated on top of two gold bond pads, which are used for the voltage measurement in a 4-terminal configuration. For each end of the nanowire, the first bond is to the nanowire with the second bond going to an empty bond pad. 

\subsection*{Low-temperature measurements}

We use 4-wire measurement circuit for measuring the mechanical resonance as shown in Fig.~\ref{fig:100nm_wire}. The sample is cooled in a dilution refrigerator with a base temperature of $\SI{7}{\milli\kelvin}$. The refrigerator is equipped with a 8\,T superconducting magnet. We sweep the frequency of the current generator at a fixed drive while measuring the voltage across the wire with a lock-in amplifier.
We found that a $\SI{50}{\nano\ampere}$ measuring current is not destructive even for the thinnest nanowires at room temperature.  Before cooling down, we test electric continuity of the nanowires using a portable digital multimeter with a $\SI{10}{\mega\ohm}$ resistor attached in series.

\subsection{Nanowires as superfluid thermometers}

Long, thin nanowires make extremely sensitive probes of  excitations in superfluid \3 \cite{Enrico95}. Pushing the superfluid  thermometry down to the ultimate, sub-\SI{100}{\micro\kelvin}, limit was the original motivation for developing the nanowires. For this use, these probes have several remarkable properties.

First, the resonance frequency $f_0$ can be made orders of magnitude lower than that of typical strained nanoresonators \cite{Golokolenov22} or even nanofabricated strain-free resonators \cite{zhou2020approaching}. That makes our devices ideal for quantum fluids research  as the non-invasive probing of quantum fluids requires operation at frequencies where the probe does not directly excite sound waves or other collective excitations \cite{Schmoranzer2011}.
%{This is not relevant information in a nanotech journal: A wire oscillating in a compressible fluid acts as a dipole emitter of sound waves, resulting in an undesired dissipation term proportional to the second power of the frequency of its oscillations\cite{Schmoranzer2011}. By choosing the frequency of the oscillator in the kilohertz range, we can bring acoustic emission to the levels much lower than the intrinsic (in vacuo) dissipation and therefore enhance the sensitivity of the probe.}

Second, the signal amplitude is proportional to resonator length. To gain the best signal to noise ratio for given velocity and magnetic field it is thus beneficial to increase the length of the wire. We have fabricated and characterised a $\SI{200}{\nano\metre}$ diameter wire of $\SI{3}{\milli\metre}$ in length with the aspect ratio of $1.5 \times 10^4$. We believe at least 1\,cm suspended length can be obtained on a suitable platform. 

Third, creating a superconducting NEMS that would  be suspended far from the substrate has been a challenging task so far \cite{Golokolenov22}. The effects of the substrate can be important even when the distance to the substrate is tens of times the NEMS thickness \cite{Gazizulin18,Guthrie2021}. In contrast, our fabrication technique allows positioning the nanowire in an arbitrary location, including across a sizable gap in the substrate. 

Finally, Euler’s model of  incompressible fluid prescribes that the velocity field of a two-dimensional flow past a sharp corner must have a singularity. In lithographically defined oscillators this results in an undesired flow enhancement near inevitable corners. Such flow fields are poorly characterised due to the uncertainty in the curvature of the edges.  For example in superfluid \3 this means that the Landau critical velocity in a flow past an object with sharp edges will be exceeded at much lower oscillator velocities than for an edgeless object \cite{Lambert1992}. The intrinsic circular cross section in our wires overcomes this problem. The models for motion of an oscillating round cylinder in a fluid are readily available and they have been thoroughly tested by decades of experimentation in quantum fluids with wires of circular cross section on the order of several micrometres\cite{tough1964,carless1983,carless1983a,Enrico95}.

%Euler’s model of  incompressible fluid prescribes that the velocity field of a two-dimensional flow past a sharp corner must have a singularity. For many measurements this is important. As an example, the energy resolution of the superfluid \3-B bolometer\cite{Bradley1995} is directly proportional to the velocity of its sensitive element -- the oscillating nanowire. Breaking the Landau velocity limit would inevitably enhance dissipation and therefore significantly reduce the sensitivity. 

%\backmatter

\section*{Contributions}

%D.E.Z. initiated this research. %Which part of the research does this refer to?
The nanowire fabrication process was developed by S.A., M.C., J.G., T.S., J.S., M.D.T., and D.E.Z. The fabrication and resonator characterization was done by S.A., M.D.T., J.R.P., and D.E.Z. The manuscript was prepared by S.A., M.D.T. and D.E.Z with contributions from all authors. S.A., A.C., R.P.H, J.M.-R., J.M., G.R.P., J.R.P., X.R., J.S., M.D.T., S.M.W., and D.E.Z acquired funding for this project.

\section*{Acknowledgements}
 We thank E. Laird and L. Levitin for stimulating discussions. This work was funded by UKRI EPSRC (EP/P024203/1, EP/X004597/1, and EP/W015730/1) and STFC (ST/T006773/1), as well as EU H2020 European Microkelvin Platform (Grant Agreement 824109). We thank `Esteves Group' for outstanding customer service.  S.A. acknowledges financial support from the Jenny and Antti Wihuri Foundation. M.D.T acknowledges financial support from the Royal Academy of Engineering (RF\textbackslash 201819\textbackslash 18\textbackslash 2).

\section*{Competing interests}
 The authors declare no competing interests.

 \section*{Data availability}
 The data that supports the findings of this study are available in Lancaster University data repository at [link/reference to be added]

 \section*{Code availability}
No specialised software was used in analysing the data.

\bibliography{sn-bibliography}

%\bibliography{sn-bibliography}% common bib file

\clearpage

\end{document}